 \documentclass[smallabstract,smallcaptions]{dccpaper}

\usepackage{epsfig}
\usepackage{cite}
\usepackage{amsmath}
\usepackage{amssymb}
\usepackage{color}
\usepackage{url}
\usepackage{algorithm}
\usepackage{algpseudocode} 
\usepackage{multirow}
\usepackage{makecell}
\usepackage{comment}
\usepackage{caption}
\usepackage[table]{xcolor}  
\definecolor{gold}{rgb}{1,0.84,0}  
\definecolor{silver}{rgb}{0.75,0.75,0.75}  
\usepackage{graphicx}
\usepackage{subcaption}
\newlength{\figurewidth}
\newlength{\smallfigurewidth}

\begin{document}

\title
{\large
\textbf{Adaptive Rate Control for Deep Video Compression with Rate-Distortion Prediction}
}

\author{%
Bowen Gu$^{1}$, Hao Chen$^{1}$, Ming Lu$^{1}$, Jie Yao$^{2}$, and Zhan Ma$^{1}$\\[0.5em]
{\small\begin{minipage}{\linewidth}\begin{center}
\begin{tabular}{ccc}
$^{1}$Nanjing University & \hspace*{0.5in} & $^{2}$T-Head Semiconductor Co., Ltd. \\
\end{tabular}
\end{center}\end{minipage}}
\thanks{The corresponding author is Hao Chen (chenhao1210@nju.edu.cn).}
}

\maketitle
\thispagestyle{empty}

\begin{abstract}
Deep video compression has made significant progress in recent years, achieving rate-distortion performance that surpasses that of traditional video compression methods. However, rate control schemes tailored for deep video compression have not been well studied. In this paper, we propose a neural network-based $\lambda$-domain rate control scheme for deep video compression, which determines the coding parameter $\lambda$ for each to-be-coded frame based on the rate-distortion-$\lambda$ (R-D-$\lambda$) relationships directly learned from uncompressed frames, achieving high rate control accuracy efficiently without the need for pre-encoding. Moreover, this content-aware scheme is able to mitigate inter-frame quality fluctuations and adapt to abrupt changes in video content. Specifically, we introduce two neural network-based predictors to estimate the relationship between bitrate and $\lambda$, as well as the relationship between distortion and $\lambda$ for each frame. Then we determine the coding parameter $\lambda$ for each frame to achieve the target bitrate. Experimental results demonstrate that our approach achieves high rate control accuracy at the mini-GOP level with low time overhead and mitigates inter-frame quality fluctuations across video content of varying resolutions.
\end{abstract}

\vspace{-0.3cm}
\section{Introduction}
\vspace{-0.2cm}
Traditional video codecs have been developed and refined for over 30 years. However, while the improvements in compression ratio have slowed down, the complexity of these conventional methods has increased significantly. As a result, further advancements within this handcrafted framework are becoming increasingly challenging to achieve, which has brought deep video codecs into the spotlight. Following the hybrid video coding pipeline, most deep video codecs \cite{8953892,10.5555/3540261.3541647,9941493,10.1145/3503161.3547845,li2023neural} tend to pursue extreme compression efficiency, without considering bitrate fluctuations or additional constraints. However, in the practical application of deep video codecs over the network, effective rate control techniques are highly desirable to enable the codecs' adaptability to underlying bandwidth fluctuations and video content transitions. 

Existing rate control works for deep video compression follow the same paradigm as traditional video compression methods. Li \textit{et al.} \cite{6849994} proposed a rate control algorithm based on the R-$\lambda$ model for H.265/High-Efficiency Video Coding (HEVC). So far, works on rate control for deep video compression have primarily focused on building an accurate relationship between bitrate and $\lambda$ and improving rate-distortion performance via reasonable bit allocation. The pioneering rate control scheme for deep video compression\cite{9746080} specially designed a parameter updating scheme, leading to a relatively low rate control accuracy.  Chen \textit{et al.} \cite{10246313} proposed a multi-pass rate control scheme, which pre-encodes the input sequence to fit rate-distortion relationships. This method is more accurate compared to \cite{9746080} but comes with significant time overhead. Zhang \textit{et al.} \cite{Zhang2024NeuralRC} proposed a one-pass rate control scheme for deep video compression based on neural networks, in which the rate implementation module doesn’t work properly when the input bitrate is out of range. 

In this paper, we propose an efficient rate control scheme for deep video compression, which leverages neural networks to predict R-$\lambda$ and D-$\lambda$ relationships without pre-encoding, ultimately determining a coding parameter $\lambda$ for each frame based on these relationships. To accomplish this, we train two predictors using a shared neural network architecture, which directly learns from uncompressed frames to estimate R-$\lambda$ or D-$\lambda$ relationships after compression for each frame. Utilizing these relationships, we introduce a fast search algorithm to identify the target distortion for each frame within a mini-GOP, and determine the optimal parameter $\lambda$ for coding accordingly.

In the design of the predictors, we introduce a distortion addition mechanism to address the potential decline in prediction performance when using uncompressed frames to fit R-$\lambda$ and D-$\lambda$ relationships. Specifically, we multiply the intermediate value of the predicted distortion from the previous frame pair by a tensor following a Gaussian random distribution, and then add it to the original reference frame of the current frame pair. This introduces a controlled amount of distortion to the reference frame, which helps to bridge the mismatch between reference frames after actual encoding and uncompressed frames. Additionally, we down-sample uncompressed frames to a fixed low resolution before inputting them into the predictors. This allows our rate control scheme to cover a wider range of video resolutions and improve computational efficiency in practical use.

To evaluate the performance of our proposed rate control scheme, we compared it against existing state-of-the-art approaches. The comparison was conducted in terms of rate error, control time consumption, and quality fluctuation, using public video datasets with different resolutions and across multiple target bitrate points. The experimental results demonstrate the efficacy of our proposed rate control scheme. It achieves high rate control accuracy with a moderate time overhead while effectively mitigating inter-frame quality fluctuations.

\textbf{Contributions.} 1) We propose an efficient rate control scheme for deep video compression, leveraging neural network-based prediction directly learned from uncompressed frames without pre-encoding. This approach ensures high rate control accuracy with moderate time overhead, effectively reducing inter-frame quality fluctuations while adapting to video content changes; 2) To address the mismatch between actual coded reference frames and uncompressed frames, we introduce a novel distortion addition mechanism that significantly enhances overall rate control performance; 3) By utilizing fixed low-resolution frames as input, we achieve consistent computational cost across videos of varying resolutions, greatly improving its practical applicability.

\vspace{-0.3cm}
\section{Related Work}
\vspace{-0.2cm}
\textbf{Deep Video Compression.}
Deep video compression techniques have largely employed a two-stage pipeline that mirrors the predictive coding architecture of traditional video codecs \cite{6849994}. DVC, as proposed by Lu \textit{et al.} \cite{8953892}, introduced an explicit methodology for encoding both motion flows and residuals within this framework. This work represented a significant step forward in applying deep learning techniques to video compression tasks. Building upon this foundation, Li \textit{et al.} \cite{10.5555/3540261.3541647} developed a more sophisticated deep contextual video compression framework. This innovative approach utilizes feature domain context as a condition, which can be flexibly designed and learned to enhance encoding, decoding, and entropy modeling processes. Subsequently, a series of deep video codecs based on this framework have been proposed \cite{9941493, 10.1145/3503161.3547845, li2023neural}, achieving superior rate-distortion performance compared to next-generation traditional codecs. 

\textbf{Rate Control for Deep Video Compression.} 
Research efforts on rate control for deep video compression have primarily concentrated on two key objectives: enhancing rate-distortion performance and ensuring rate control accuracy. Li \textit{et al.} \cite{9746080} pioneered rate control for deep video compression. This method highly relies on coding experiences, limiting its accuracy in situations where video content changes rapidly. Building on this, Chen \textit{et al.} \cite{10246313} proposed a simple yet effective rate control algorithm for end-to-end video coding by introducing a rescale ratio and a generalized R-D model, which converts sparsely distributed R-D points to denser points without introducing additional models. However, the pre-encoding step results in significant time consumption. Zhang \textit{et al.} \cite{Zhang2024NeuralRC} introduced the first neural network-based rate control scheme specifically designed for deep video codecs, significantly improving rate-distortion performance. However, their rate implementation module faces challenges when the input bitrate falls outside the expected range. Xu \textit{et al.} \cite{bit allocation} introduced an innovative paradigm of bit allocation within latent domain, which can serve as an empirical bound on the R-D performance of bit allocation. However, it comes at the cost of substantial computational resources. Existing works mentioned above only achieve GOP-level rate control and incur significant time overhead, limiting their applicability in real-time scenarios.

\vspace{-0.3cm}
\section{Method}
\vspace{-0.2cm}

\subsection{System Framework}
\vspace{-0.1cm}
To achieve rate control for deep video compression, it is necessary to establish a relationship between the bitrate R and the coding parameter $\lambda$. The mainstream rate control algorithm in H.265/HEVC adopts the hyperbolic law to describe the bitrate versus $\lambda$ relationship. Previous studies \cite{10246313,9810725} further prove that deep codecs share similar rate-distortion characteristics with traditional codecs. Therefore, we adopt the hyperbolic law in this paper to describe the R-$\lambda$ relationship and D-$\lambda$ relationship. Experimental results in Sec. \ref{ssec:results} further demonstrate the accuracy of the hyperbolic law.
In detail, the relationships can be described in the following way:
\begin{equation}
\setlength{\abovedisplayskip}{2pt}
\setlength{\belowdisplayskip}{3pt}
D(R)=C R^{-K}.
\label{RDcurve}
\end{equation}$
\lambda$ is the slope of the R-D curve, i.e., \begin{equation}
\setlength{\abovedisplayskip}{2pt}
\setlength{\belowdisplayskip}{3pt}
\lambda=-\frac{\partial D}{\partial R}.
\label{D/R}
\end{equation}
Taking Eq. \eqref{RDcurve} into Eq. \eqref{D/R} yields
\begin{equation}
\setlength{\abovedisplayskip}{2pt}
\setlength{\belowdisplayskip}{3pt}
R = \alpha_1 \lambda^{\beta_1}.
\label{Rlam}
\end{equation}
Similarly, we can easily derive the relationship between distortion and $\lambda$: 
\begin{equation}
\setlength{\abovedisplayskip}{2pt}
\setlength{\belowdisplayskip}{3pt}
D = \alpha_2 \lambda^{\beta_2}.
\label{Dlam}
\end{equation}
Under this premise, the most intuitive approach to determining the hyper-parameters is to encode the frames with different $\lambda$s and get a set of (R,$\lambda$) or (D,$\lambda$) points to fit the relationships, which is so-called a multi-pass way. However, this approach is rather time-consuming, which is unacceptable in scenarios such as live streaming and real-time communication. The rate control schemes in \cite{6849994,9746080} reduce time consumption and improve rate control accuracy by proposing a parameter updating scheme. 

\begin{figure}[t]
    \centering
    \setlength{\abovecaptionskip}{0cm}
    \includegraphics[width=.7\textwidth]{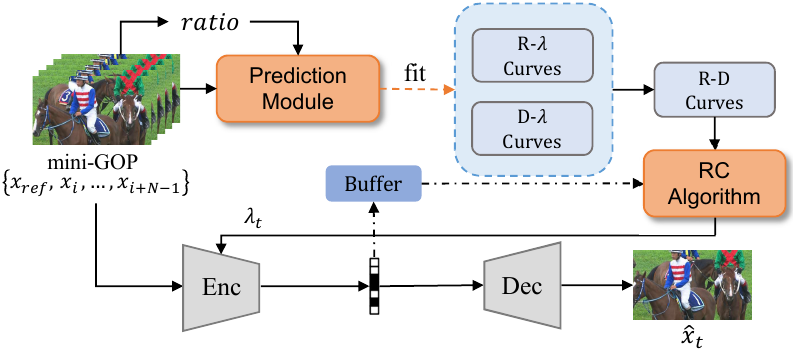}  
    \caption{The system framework of our rate control scheme.}
    \label{fig:workflow}
    \vspace{-0.5cm}
\end{figure}

Our method employs neural networks to predict R-$\lambda$ and D-$\lambda$ relationships in a one-shot manner, providing both accuracy and efficiency. Fig.~\ref{fig:workflow} shows the framework of our rate control scheme. We input all adjacent frame pairs within a mini-GOP into the prediction module (Sec. \ref{ssec:rd_prediction}). The output gives us samples of (R,$\lambda$) points and (D,$\lambda$) points, allowing us to fit R-$\lambda$ and D-$\lambda$ relationships of each frame with the least squares method. This prediction module enables content-adaptive rate control, preserving high accuracy even when the video content undergoes significant changes, in contrast to the limitations of previous methods. Given the target bitrate, we then search for the optimal bit allocation combination of frames within the mini-GOP which alleviates quality fluctuations using our rate control (RC) algorithm (Sec. \ref{ssec:rate_control}). The rate control algorithm takes the predicted relationships for each frame as input, and formulates the bit allocation as an optimization problem. Via reasonable bit allocation within the mini-GOP, our rate control algorithm can effectively smooth out quality fluctuations and ensure that the overall bitrate constraint is satisfied. Finally, we start the actual coding process using the derived coding parameters, and dynamically update the bit allocation as needed to maintain the output bitrate. 

By leveraging predicted R-$\lambda$ and D-$\lambda$ relationships instead of conducting actual encoding, our proposed rate control scheme synthesizes the strengths of both multi-pass and one-pass methodologies. This novel approach presents a solution that achieves high control accuracy while maintaining time efficiency, addressing the longstanding trade-off between precision and speed in rate control for deep video compression. 

\vspace{-0.1cm}
\subsection{Prediction Module}
\label{ssec:rd_prediction}
\vspace{-0.1cm}

\begin{figure}[t]
    \centering
    \setlength{\abovecaptionskip}{0.2cm}
    \includegraphics[width=.8\textwidth]{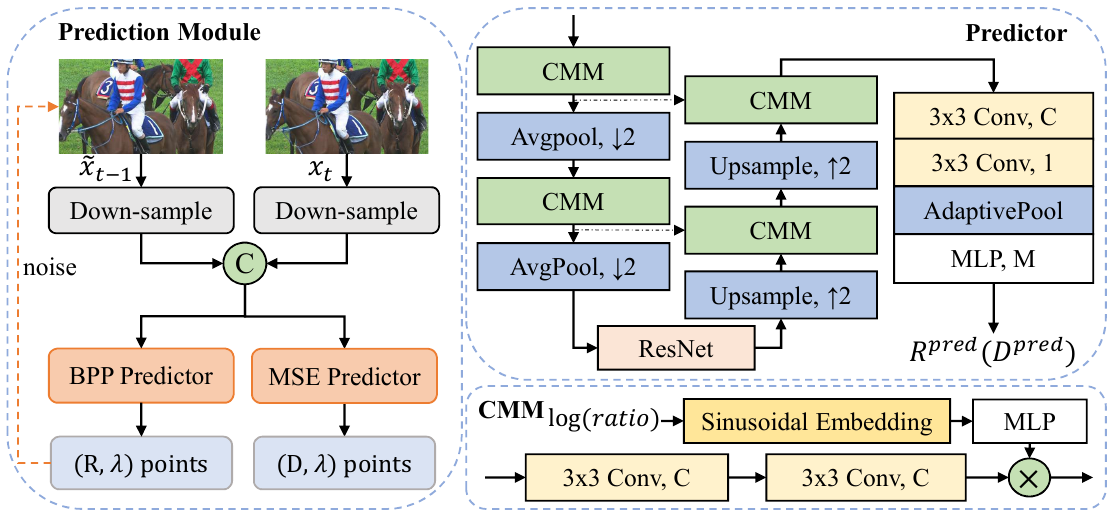}  
    \caption{Network structure of our prediction module.}
    \label{fig:network}
    \vspace{-0.5cm}
\end{figure}

The prediction module in Fig.~\ref{fig:network} comprises a BPP predictor (bits per pixel) and an MSE predictor (mean squared error), sharing the same network structure. These predictors are tasked with predicting (R,$\lambda$) and (D,$\lambda$) points for each frame, respectively.

We first down-sample the current frame and its reference frame to 240P resolution. The input feature of predictors is the channel-wise concatenation of the down-sampled frame pair. In the CMM (Conv-Multiply-MLP) module, the logarithmically scaled down-sample ratio is sinusoidally embedded to encode it into a fixed-dimensional space, and then passed through an MLP layer. The embedding operation enables the network to better perceive the original resolution of the input frames, thereby enhancing robustness to varying resolutions. $C$ is the number of output channels of the convolutional layer and $M$ is the size of a predefined $\lambda$ set. To reduce computational complexity while extracting sufficient information from the video frames, we design a symmetric structure that down-samples the input feature several times with convolutional layers and pooling layers to explore deep features. After down-sampling operations, we restore the deep feature to the original size by several up-sampling operations. We also aggregate multi-level features by concatenating features of the same level. Finally, the top-level aggregated feature is fed into two convolutional layers and an MLP layer to output the predicted bitrate $\text{R}^{pred}=\{bpp_1^{pred}, \ldots, bpp_M^{pred}\}$ or distortion $\text {D}^{pred}=\{mse_1^{pred}, \ldots, mse_M^{pred}\}$ under the predefined $\lambda$ set. It is important to note that the $\lambda$ set should include both the maximum and minimum values of the available $\lambda$s to determine whether the target bitrate can be achieved. 

The loss functions for our prediction modules are: 
\begin{equation}
\setlength{\abovedisplayskip}{2pt}
\setlength{\belowdisplayskip}{3pt}
L_{R} = \frac{1}{M}\sum_{i=0}^{M-1}|bpp^{pred}_i-bpp^{real}_i|, \quad
L_{D} = \frac{1}{M}\sum_{i=0}^{M-1}|mse^{pred}_i-mse^{real}_i|,
\label{loss}
\end{equation}
where $bpp^{real}_i$ and $mse^{real}_i$ are the actual output bitrate and distortion, respectively, after encoding with the $i^{\text{th}}$ coding parameter from the predefined $\lambda$ set.

We found that directly inputting uncompressed frames led to a decline in prediction performance. This is because the R-$\lambda$ and D-$\lambda$ relationships for each frame depend on the current frame to be coded and the already coded reference frame. To address this issue, we propose a distortion addition mechanism. Specifically, we multiply the predicted distortion from the previous frame pair by a tensor following a Gaussian random distribution and add it to the uncompressed reference frame of the current frame pair. The distortion to be added can be expressed as:
\begin{equation}
\setlength{\abovedisplayskip}{2pt}
\setlength{\belowdisplayskip}{3pt}
D_{add}=T\cdot \sqrt{\ln \frac{e^{D_{max}}+e^{D_{min}}}{2}},  \quad \text{where} \quad T \sim \mathcal{N}(0, 1),
\label{noise}
\end{equation}
where $D_{max}$ and $D_{min}$ are the maximum value and minimum value of the predicted distortion (in the form of MSE) of the last frame pair, and $T$ is a tensor with the same shape as the input frame, with elements randomly generated from a standard Gaussian distribution. This simulates the distortion introduced during the coding process, improving the prediction accuracy. 

\vspace{-0.1cm}
\subsection{Rate Control Algorithm}
\label{ssec:rate_control}
\vspace{-0.1cm}

\setlength{\textfloatsep}{8pt}
\begin{algorithm}[t]
\caption{$\lambda$-domain rate control}
\begin{algorithmic}[1]
\Statex \textbf{Input:} Target bitrate $R_{tar}$, predicted R-$\lambda$ and R-D relationships of each frame, set of the minimum value of predicted distortion of each frame $\{mse^{min}_0, mse^{min}_1, ..., mse^{min}_{N-1}\}$, set of the maximum value of predicted distortion of each frame $\{mse^{max}_0, mse^{max}_1, ..., mse^{max}_{N-1}\}$.
\Statex \textbf{Output:} Coding parameters $\lambda_i, i=0,1,...,N-1$, for each frame in a mini-GOP.
\State ${D}_{\text{LB}} \leftarrow \max \{mse^{min}_0, mse^{min}_1, ..., mse^{min}_{N-1}\}$
\Statex ${D}_{\text{UB}} \leftarrow \min \{mse^{max}_0, mse^{max}_1, ..., mse^{max}_{N-1}\}$
\State \textbf{for} $count=0$ to $K$ \textbf{do} 
\State\quad ${D}_{tar} \leftarrow \frac{{D}_{\text{LB}} + {D}_{\text{UB}}}{2}$
\State\quad Determine $R_{i}, i=0,1,...,N-1$, with Eq.~\eqref{RDcurve}, setting $D$ as ${D}_{tar}$
\State\quad $R_{total} \leftarrow \sum_{i=0}^{N-1} R_{i}$
\State\quad\textbf{if} $\left| \frac{R_{total}-R_{tar}}{R_{tar}}\right| < \epsilon$  \textbf{then}
\State\quad\quad \textbf{break}
\State\quad \textbf{else if} $R_{total} < R_{tar}$ \textbf{then}
\State\quad\quad ${D}_{\text{UB}}\leftarrow{D}_{tar}$
\State\quad \textbf{else}
\State\quad\quad ${D}_{\text{LB}}\leftarrow{D}_{tar}$
\State Derive bit ratios $r_{i} = \frac{R_{i}}{\sum_{j=0}^{N-1} R_{j}}, i=0,1,...,N-1$
\State Initialize $buffer = 0$, $ratio_{sum} = \sum_{i=0}^{N-1} r_{i}$
\State\textbf{for} frame $i = 0 $ to $N$ in a mini-GOP \textbf{do}
\State\quad $R_i\leftarrow R_{tar} \cdot \frac{r_{i}}{ ratio_{sum}} + buffer$
\State\quad Determine $\lambda_i$ with Eq.~\eqref{Rlam} and start encoding
\State\quad Put the surplus or deficit bits into the buffer
\State\textbf{end for}
\end{algorithmic}
\label{alg:lambda_domain_rate_control}
\end{algorithm}

In deep video compression, significant quality fluctuations often occur during the initial frames of a GOP, particularly within the first mini-GOP. The rate control scheme proposed in \cite{10246313} addresses this issue by smoothing quality fluctuations between frames using rescale ratios. However, no $\lambda$-domain rate control scheme for variable-rate deep video codecs has yet been explored to achieve improved quality consistency. To address this gap, we propose a novel rate control algorithm designed to minimize quality fluctuations while adhering to target bitrate constraints. Upon R-D prediction, our algorithm manages cases where the target bitrate exceeds the codec's bitrate range. Specifically, when the target bitrate surpasses the maximum bitrate limit, we encode using the highest $\lambda$ value; conversely, when the target bitrate falls below the minimum, we use the lowest $\lambda$ value. This mechanism enables the codec to output a bitstream closest to the unachievable target bitrate within the bitrate range, a capability that the method proposed in \cite{Zhang2024NeuralRC} does not provide.
If the target bitrate is within the acceptable range, the rate control algorithm proceeds as normal. As outlined in Algorithm \ref{alg:lambda_domain_rate_control}, the algorithm begins by establishing the upper and lower bounds of the target distortion achievable across all frames within a mini-GOP. Subsequently, we employ an iterative binary search algorithm to approximate the optimal target distortion for all frames in a mini-GOP, with the objective of minimizing the rate error between the target bitrate $R_{tar}$ and the total bitrate $R_{total}$. This iterative process is meticulously designed to identify bit allocation ratios that concurrently minimize inter-frame quality fluctuations and maintain high rate control accuracy. Finally, the optimal coding parameter $\lambda$ is calculated for each frame, and the coding process is initiated.

\vspace{-0.3cm}
\section{Experiments}
\label{sec:experiments}
\vspace{-0.2cm}

\subsection{Experimental Setup}
\label{ssec:setup}
\vspace{-0.1cm}

\noindent\textbf{Datasets:} 
The training dataset for our prediction module comprised original full-resolution video sequences obtained from Vimeo (\url{https://vimeo.com/}). To enhance prediction accuracy, we augmented this dataset with video sequences from YouHQ \cite{zhou2024upscale}. We used the HEVC B, C, D, and E datasets for testing, encompassing 16 video sequences with resolutions ranging from $1920\times1080$ to $416\times240$. Additionally, we evaluated our scheme's performance on the validation set of Vimeo-90K septuplet dataset, which includes over 7000 short video clips.

\noindent\textbf{Implementation:} We implemented variable-rate deep video codecs based on DVC\cite{9506269} and DCVC-HEM \cite{10.1145/3503161.3547845}. During each iteration of the training for the prediction module, we randomly cropped each 5-frame clip into patches of varying size: $1920\times1088$, $1280\times768$, $832\times512$, and $448\times256$. Each frame was initially coded with the codec under each $\lambda$ in the predefined $\lambda$ set, with the resultant values (i.e., $bpp^{real}_i$ or $mse^{real}_i$) concatenated to form $R^{real}$ or $D^{real}$ as training labels. Subsequently, one of the output reconstructed images was randomly selected as the reference frame for the subsequent frame to be coded. The initial learning rate was set to 1e-4 and was reduced by a factor of $0.1$ when the performance on the validation set failed to improve for four consecutive epochs. The size of a mini-GOP ($N$) was set to 4, while $M$ was fixed at 8. The $\lambda$ set was defined as 8 exponentially interpolated values between the minimum and maximum values of $\lambda$. In Algorithm \ref{alg:lambda_domain_rate_control}, the parameters $K$ and $\epsilon$ were set to 100 and 0.01, respectively. 

\noindent\textbf{Baselines:} We benchmarked our rate control scheme against both multi-pass and one-pass approaches. The multi-pass scheme is based on the full resolution rate control scheme in \cite{10246313}, where each frame is coded with multiple passes with parameters in the predefined $\lambda$ set. The one-pass scheme is based on \cite{6849994} and \cite{9746080}. Both baseline methods incorporate parameter initialization, bit allocation, and parameter updating.

\noindent\textbf{Metrics:} We evaluated our rate control scheme at three bitrate points, using the average as the result. Rate control accuracy was quantified using the relative bitrate error: $\Delta \mathrm{R} = \left| \frac{R_{\text{real}} - R_{\text{tar}}}{R_{\text{tar}}} \right|$. Our method implements rate control and calculates relative bitrate error at the mini-GOP level, whereas baseline approaches operate at the GOP level. This finer granularity makes achieving high accuracy in our rate control more challenging. Rate control efficiency was assessed using the rate control time: $T_{RC} = \frac{\text{Rate Control Time}}{\text{Encoding Time}}$. For a fair comparison, we excluded bit allocation time when calculating $T_{RC}$ for multi-pass and one-pass approaches. Quality fluctuation was defined as $Q_{F} = \frac{\frac{1}{N} \sum_{i=0}^{N-1} | MSE_i - \mu |}{\mu}$, where $\mu = \frac{1}{N} \sum_{i=0}^{N-1} MSE_i$. The fluctuation ratio was calculated as the ratio of $Q_{F}$ after rate control to that of fixed $\lambda$ coding.

\vspace{-0.1cm}
\subsection{Experimental Results}
\label{ssec:results}
\vspace{-0.1cm}

\begin{table*}[t]
\setlength{\abovecaptionskip}{0.1cm}
\caption{Comparison of Rate Control Accuracy and Efficiency. Data marked in red and blue indicate the best and the second best performance respectively.}
\renewcommand{\arraystretch}{1.2}
\resizebox{\textwidth}{!}{%
\setlength{\tabcolsep}{2pt} 
\begin{tabular}{c|c|cccccccccc|cc}
\Xhline{1.5pt} 
\multirow{2}{*}{\large{Codec}} & \multirow{2}{*}{\large{Method}} &\multicolumn{2}{c}{\large{HEVC B}} & \multicolumn{2}{c}{\large{HEVC C}} & \multicolumn{2}{c}{\large{HEVC D}} & \multicolumn{2}{c}{\large{HEVC E}} & \multicolumn{2}{c|}{\large{Vimeo test}} & \multicolumn{2}{c}{\large{Average}} \\
& &\large{$\Delta \mathrm{R}(\%)$} & \large{$T_{RC}$} & \large{$\Delta \mathrm{R}(\%)$} & \large{$T_{RC}$} & \large{$\Delta \mathrm{R}(\%)$} & \large{$T_{RC}$} & \large{$\Delta \mathrm{R}(\%)$} & \large{$T_{RC}$} & \large{$\Delta \mathrm{R}(\%)$} & \large{$T_{RC}$} & \large{$\Delta \mathrm{R}(\%)$} & \large{$T_{RC}$}\\
\hline 
\multirow{2}{*}{\normalsize{DVC}} & Ours & \textcolor{blue}{\large8.60} & \textcolor{blue}{\large1.19} & \textcolor{blue}{\large2.80} & \textcolor{blue}{\large1.06} & \textcolor{blue}{\large3.11} & \textcolor{blue}{\large0.83} & \textcolor{blue}{\large4.94} & \textcolor{blue}{\large1.11} & \textcolor{blue}{\large2.81} & \textcolor{blue}{\large1.08} & \textcolor{blue}{\large4.45} & \textcolor{blue}{\large1.05}\\
\multirow{2}{*}{\large{\cite{9506269}}} &\large{multi-pass}& \textcolor{red}{\large3.40} & \large3.01 & \textcolor{red}{\large1.25} & \large3.06 & \textcolor{red}{\large1.69} & \large3.01 & \textcolor{red}{\large1.51} & \large3.09 & \textcolor{red}{\large1.70} & \large2.21 & \textcolor{red}{\large1.91} &\large2.88\\
& \large{one-pass} & \large13.45 & \textcolor{red}{\large0.01} & \large3.67 & \textcolor{red}{\large0.01} & \large7.58 & \textcolor{red}{\large0.01} & \large12.32 & \textcolor{red}{\large0.01} & \large8.94 & \textcolor{red}{\large0.01} &\large9.19 & \textcolor{red}{\large0.01}\\
\hline
\multirow{2}{*}{\normalsize{DCVC-HEM}}& Ours & \textcolor{red}{\large2.27} & \textcolor{blue}{\large0.58} & \textcolor{red}{\large4.39} & \textcolor{blue}{\large0.55} & \textcolor{red}{\large2.85} & \textcolor{blue}{\large0.49} & \textcolor{red}{\large1.55} & \textcolor{blue}{\large0.58} & \textcolor{red}{\large5.25} & \textcolor{blue}{\large0.75} & \textcolor{red}{\large3.26} & \textcolor{blue}{\large0.59}\\
\multirow{2}{*}{\large{\cite{10.1145/3503161.3547845}}} &\large{multi-pass}& \textcolor{blue}{\large9.21} & \large1.99 & \textcolor{blue}{\large6.21} & \large1.99 & \textcolor{blue}{\large8.06} & \large2.02 & \textcolor{blue}{\large9.66} & \large1.99 & \textcolor{blue}{\large7.54} & \large1.73 & \textcolor{blue}{\large8.14} &\large1.94\\
& \large{one-pass} & \large11.64 & \textcolor{red}{\large0.01} & \large9.78 & \textcolor{red}{\large0.01} & \large12.84 & \textcolor{red}{\large0.01} & \large13.46 & \textcolor{red}{\large0.01} & \large9.84 & \textcolor{red}{\large0.01} &\large11.51 &\textcolor{red}{\large0.01}\\
\Xhline{1.5pt} 
\end{tabular}
}
\label{tab:performance}
\end{table*}

\begin{table*}[t]
\setlength{\abovecaptionskip}{0.1cm}
\centering
\caption{Comparison of Fluctuation Ratio (\%).}
\renewcommand{\arraystretch}{0.9}
\setlength{\tabcolsep}{3pt} 
\scalebox{0.9}{
\begin{tabular}{c|c|ccccc|cc}
\Xhline{1.5pt} 
\small{Codec}& \small{Method}& \small{HEVC B} & \small{HEVC C} & \small{HEVC D} & \small{HEVC E} & \small{Vimeo test} & \small{Average} \\
\hline 
& \small{multi-pass} & \small232.70 & \small113.78 & \small114.21 & \small114.36 & \small126.10 & \small140.23\\
\small{DVC} & \small{one-pass} & \small237.04 & \small165.51 & \small139.76 & \small120.76 & \small165.51 & \small165.72\\
& \small{Ours} & \textbf{\small99.12} & \textbf{\small82.12} & \textbf{\small84.65} & \textbf{\small81.02} & \textbf{\small74.20} & \textbf{\small84.22}\\
\hline
& \small{multi-pass} & \small223.34 & \small125.42 & \small122.30 & \small193.55 & \small155.51 &\small164.02 \\
\small{DCVC-HEM} & \small{one-pass} & \small116.09 & \small101.80 & \small101.79 & \small130.42 & \small141.56 & \small118.33\\
& \small{Ours} & \textbf{\small70.83} & \textbf{\small82.79} & \textbf{\small75.42} & \textbf{\small36.27} & \textbf{\small95.19}  & \textbf{\small72.10}\\
\Xhline{1.5pt} 
\end{tabular}
}
\label{tab:qualityfluctuation}
\end{table*}

\begin{figure}[t]
    \centering
    \setlength{\abovecaptionskip}{0.1cm}
    \begin{subfigure}[b]{0.45\textwidth} 
         \setlength{\abovecaptionskip}{0.1cm}
        \includegraphics[width=\textwidth]{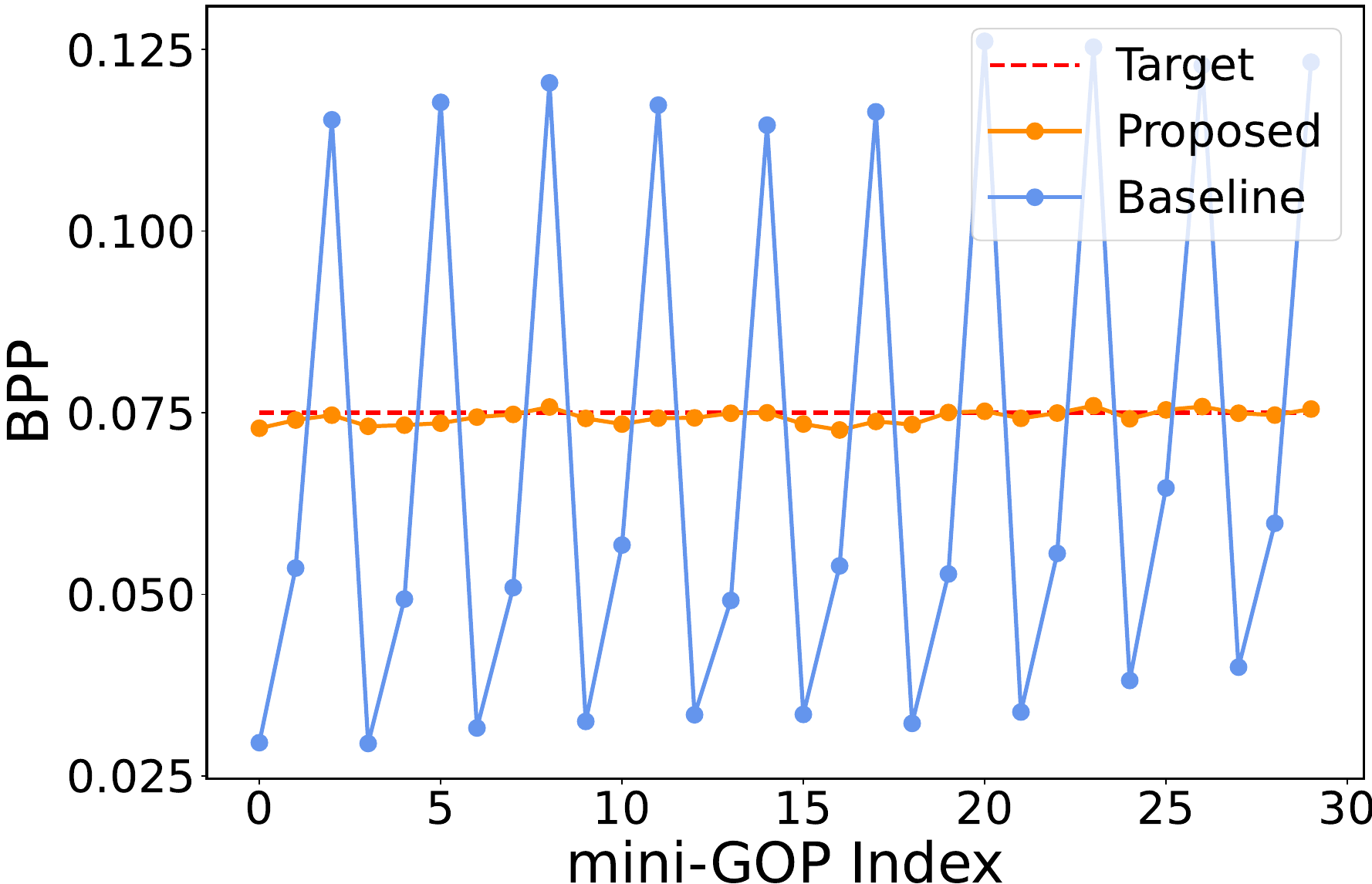} 
        \caption{Rate control accuracy} 
        \label{fig:image1}
    \end{subfigure}
    \hfill
    \begin{subfigure}[b]{0.45\textwidth} 
        \setlength{\abovecaptionskip}{0.1cm}
        \includegraphics[width=\textwidth]{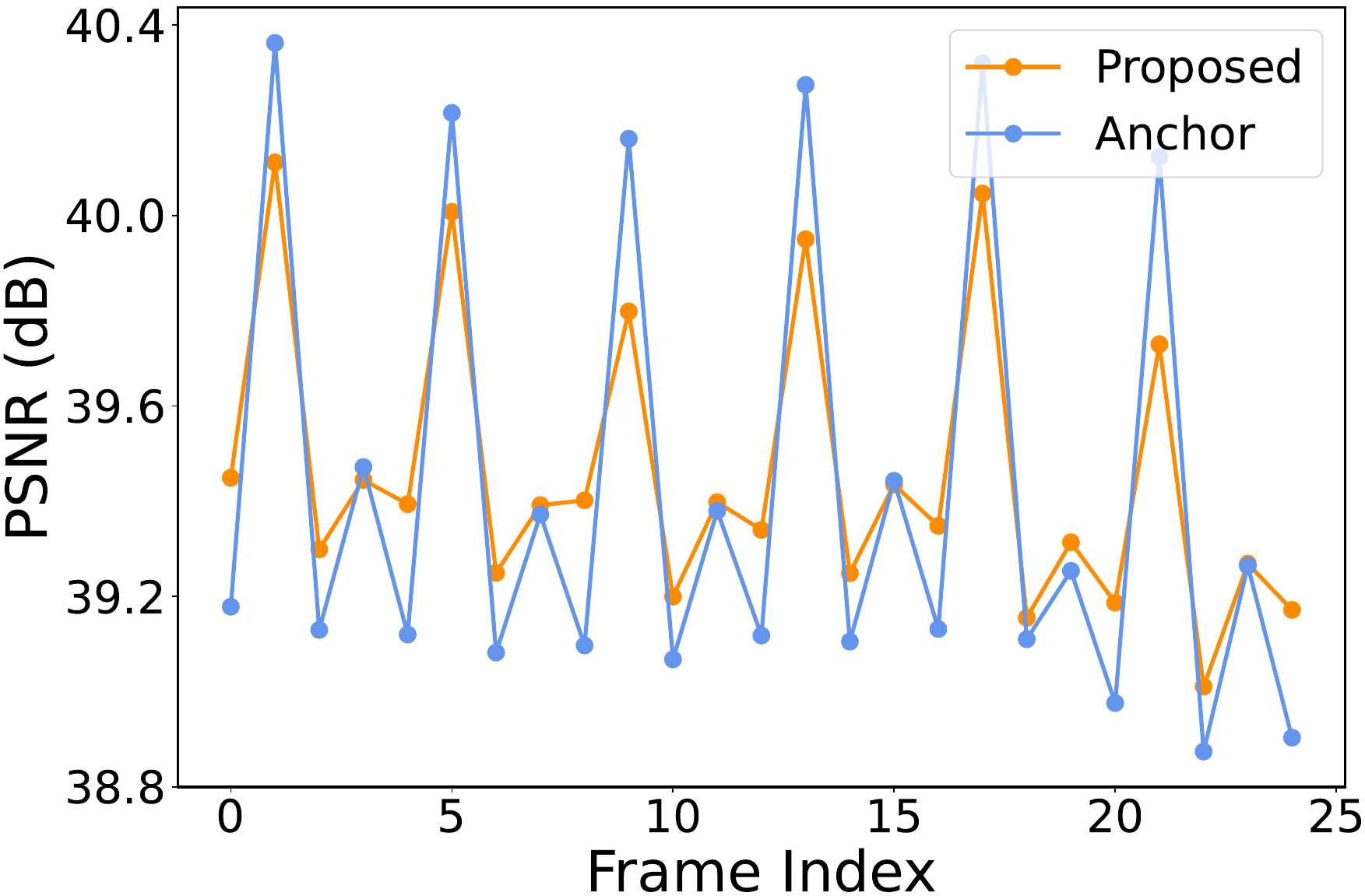} 
        \caption{Quality fluctuations} 
        \label{fig:image2}
    \end{subfigure}
    \caption{Rate control results on HEVC class E FourPeople sequence. The target bpp is set as 0.075. ``Baseline'' denotes the multi-pass rate control method, and ``Anchor'' denotes the fixed $\lambda$ coding approach.}  
    \label{fig:two_images}
\end{figure}

\noindent\textbf{Rate Control Accuracy and Efficiency:} As shown in Tab. \ref{tab:performance}, our proposed rate control scheme demonstrates superior performance across all test sequences. It consistently achieves either the minimum or second-minimum bitrate error while maintaining an acceptably low time complexity. When employing DVC as the codec, the bitrate error ($\Delta \mathrm{R}$) using our approach is marginally higher than that of the multi-pass approach, with a difference of 2.54\%. However, with DCVC-HEM as the codec, the bitrate error is further reduced, surpassing the performance of the multi-pass method. This enhanced performance can be attributed to the narrower bitrate range of DCVC-HEM compared to DVC. Our rate control scheme meticulously accounts for the bitrate range of each frame during allocation, thereby minimizing bitrate error. In contrast, the multi-pass method may allocate bits beyond the acceptable range for individual frames, resulting in larger bitrate errors. 
Fig. \ref{fig:image1} further illustrates a comparative analysis of rate control accuracy at the mini-GOP level. As shown, our proposed approach maintains precise rate control. In contrast, the multi-pass approach exhibits notable deviations from the target bitrate. 
Furthermore, our scheme exhibits a substantially lower rate control time ($T_{RC}$) compared to the multi-pass approach, accounting for only 36.5\% and 30.4\% of the time for DVC and DCVC-HEM, respectively. Notably, when using DCVC-HEM, $T_{RC}$ is approximately half of that observed when using DVC. This efficiency gain is due to DCVC-HEM's slower coding speed, while our prediction module's speed remains independent of the coding speed. Consequently, our method becomes increasingly efficient with slower codecs. Notably, the one-pass approach relies solely on empirical models for rate control, resulting in minimal time overhead. However, this comes at the expense of low accuracy and poor robustness to diverse video content.

\noindent\textbf{Quality Fluctuation:} 
Tab. \ref{tab:qualityfluctuation} demonstrates the effectiveness of our rate control scheme in reducing quality fluctuations. Given that significant quality fluctuations usually occur in the first mini-GOP, we only calculate the fluctuation ratio in every first mini-GOP of each sequence. Our rate control scheme reduces quality fluctuations across all test sequences, whereas both multi-pass and one-pass approaches significantly increase quality fluctuations after applying rate control. Fig. \ref{fig:image2} provides an example of quality fluctuations in the first mini-GOP using our rate control scheme, where we set the output bitrate of fixed $\lambda$ coding as the target bitrate. The results clearly demonstrate that, compared to fixed $\lambda$ coding, the proposed method significantly reduces quality fluctuations while maintaining a comparable overall quality level. This reduction in frame-to-frame quality fluctuations is achieved without compromising the average video quality. 

\noindent Additional experimental results from ablation studies can be found at \url{https://github.com/NJUVISION/AdaptiveRC}.

\vspace{-0.3cm}
\section{Conclusion}
In this paper, we present a one-pass neural network-based $\lambda$-domain rate control scheme for deep video compression. Our approach introduces an efficient prediction module for content-aware prediction of the R-D-$\lambda$ relationships for frames to be coded, which helps mitigate inter-frame quality fluctuations and maintain high rate control accuracy with our rate control algorithm. We evaluate our method across two deep video codecs, varying resolutions, and three target bitrates. Experimental results demonstrate the effectiveness of our scheme in improving accuracy and reducing quality fluctuations. Future work will focus on enhancing rate-distortion performance and mitigating quality fluctuations with a more lightweight network.
\vspace{-0.2cm}

\vspace{-0.2cm}
\section{Acknowledgment}
This work was supported in part by Natural Science Foundation of China under Grant No.62471215, 62401251, and 62231002, and in part by Jiangsu Provincial Key Research and Development Program under Grant No.BE2022155. The authors would like to express their sincere gratitude to the Interdisciplinary Research Center for Future Intelligent Chips (Chip-X) and Yachen Foundation for their invaluable support.
\vspace{-0.1cm}

\vspace{-0.1cm}
\Section{References}

\begin{thebibliography}{10}

\bibitem{8953892}
Guo Lu, Wanli Ouyang, Dong Xu, Xiaoyun Zhang, Chunlei Cai, and Zhiyong Gao,
\newblock ``Dvc: An end-to-end deep video compression framework,''
\newblock in {\em 2019 IEEE/CVF Conference on Computer Vision and Pattern Recognition (CVPR)}, 2019, pp. 10998--11007.

\bibitem{10.5555/3540261.3541647}
Jiahao Li, Bin Li, and Yan Lu,
\newblock ``Deep contextual video compression,''
\newblock in {\em Proceedings of the 35th International Conference on Neural Information Processing Systems}, Red Hook, NY, USA, 2024, NIPS '21, Curran Associates Inc.

\bibitem{9941493}
Xihua Sheng, Jiahao Li, Bin Li, Li~Li, Dong Liu, and Yan Lu,
\newblock ``Temporal context mining for learned video compression,''
\newblock {\em IEEE Transactions on Multimedia}, vol. 25, pp. 7311--7322, 2023.

\bibitem{10.1145/3503161.3547845}
Jiahao Li, Bin Li, and Yan Lu,
\newblock ``Hybrid spatial-temporal entropy modelling for neural video compression,''
\newblock in {\em Proceedings of the 30th ACM International Conference on Multimedia}, New York, NY, USA, 2022, MM '22, p. 1503–1511, Association for Computing Machinery.

\bibitem{li2023neural}
Jiahao Li, Bin Li, and Yan Lu,
\newblock ``Neural video compression with diverse contexts,''
\newblock in {\em Proceedings of the IEEE/CVF Conference on Computer Vision and Pattern Recognition}, 2023, pp. 22616--22626.

\bibitem{6849994}
Bin Li, Houqiang Li, Li~Li, and Jinlei Zhang,
\newblock ``$\lambda $ domain rate control algorithm for high efficiency video coding,''
\newblock {\em IEEE Transactions on Image Processing}, vol. 23, no. 9, pp. 3841--3854, 2014.

\bibitem{9746080}
Yanghao Li, Xinyao Chen, Jisheng Li, Jiangtao Wen, Yuxing Han, Shan Liu, and Xiaozhong Xu,
\newblock ``Rate control for learned video compression,''
\newblock in {\em ICASSP 2022 - 2022 IEEE International Conference on Acoustics, Speech and Signal Processing (ICASSP)}, 2022, pp. 2829--2833.

\bibitem{10246313}
Jiancong Chen, Meng Wang, Pingping Zhang, Shurun Wang, and Shiqi Wang,
\newblock ``Sparse-to-dense: High efficiency rate control for end-to-end scale-adaptive video coding,''
\newblock {\em IEEE Transactions on Circuits and Systems for Video Technology}, vol. 34, no. 5, pp. 4027--4039, 2024.

\bibitem{Zhang2024NeuralRC}
Yiwei Zhang, Guo Lu, Yunuo Chen, Shen Wang, Yibo Shi, Jing Wang, and Li~Song,
\newblock ``Neural rate control for learned video compression,''
\newblock in {\em International Conference on Learning Representations}, 2024.

\bibitem{bit allocation}
Tongda Xu, Han Gao, Chenjian Gao, Yuanyuan Wang, Dailan He, Jinyong Pi, Jixiang Luo, Ziyu Zhu, Mao Ye, Hongwei Qin, Yan Wang, Jingjing Liu, and Ya-Qin Zhang,
\newblock “Bit allocation using optimization,”
\newblock in {\em Proceedings of the 40th International Conference on Machine Learning}, 2023, pp. 38377--38399.

\bibitem{9810725}
Chuanmin Jia, Ziqing Ge, Shanshe Wang, Siwei Ma, and Wen Gao,
\newblock ``Rate distortion characteristic modeling for neural image compression,''
\newblock in {\em 2022 Data Compression Conference (DCC)}, 2022, pp. 202--211.

\bibitem{zhou2024upscale}
Shangchen Zhou, Peiqing Yang, Jianyi Wang, Yihang Luo, and Chen~Change Loy,
\newblock ``Upscale-a-video: Temporal-consistent diffusion model for real-world video super-resolution,''
\newblock in {\em Proceedings of the IEEE/CVF Conference on Computer Vision and Pattern Recognition}, 2024, pp. 2535--2545.

\bibitem{9506269}
Jianping Lin, Dong Liu, Jie Liang, Houqiang Li, and Feng Wu,
\newblock ``A deeply modulated scheme for variable-rate video compression,''
\newblock in {\em 2021 IEEE International Conference on Image Processing (ICIP)}, 2021, pp. 3722--3726.


\end{thebibliography}

\end{document}